\documentclass[journal]{new-aiaa}
\usepackage[utf8]{inputenc}

\usepackage{graphicx}
\usepackage{amsmath}
\usepackage[version=4]{mhchem}
\usepackage{siunitx}
\usepackage{subfig}
\usepackage{longtable,tabularx}
\title{Construction of J2-Invariant Periodic Relative Motion in Highly Elliptical Orbits}

\author{Jackson Kulik\footnote{Corresponding author-jackson.kulik@ttu.edu}}
\affil{Department of Mathematics and Statistics, Texas Tech University, Lubbock, TX 79409, USA}

\begin{document}

\maketitle

\begin{abstract}
Two satellites with mean orbital elements which differ only in terms of right ascension of the ascending node, argument of perigee, and mean anomaly are notable for having the same mean orbital element secular drift rates due to the J2 perturbation. The relative orbits which result from this configuration are discounted in the literature for not providing sufficiently many degrees of freedom with which to design relative orbit geometries suitable for real world missions. However, this paper will explore a variety of useful geometries which result from this limited design space, and provide analytical formulas for the differences between chief and deputy mean orbital elements as a function of the desired orbit geometry and inertial chief orbit. These include in-track-centered, quadrant-time-centered, offset-circular, rectilinear, boomerang-shaped, and cross-track only geometries.
\end{abstract}

\section*{Nomenclature}

{\renewcommand\arraystretch{1.0}
\noindent\begin{longtable*}{@{}l @{\quad=\quad} l@{}}
$a$  & semi-major axis (km) \\
$e$ &    eccentricity (unitless)\\
$i$ & inclination (degrees in figures/radians elsewhere) \\
$\Omega$ & right ascension of the ascending node (RAAN) (degrees in figures/radians elsewhere) \\
$\omega$ & argument of perigee (degrees in figures/radians elsewhere)\\
$M$ & mean anomaly (degrees in figures/radians elsewhere) \\
$\nu$ & true anomaly (degrees in figures/radians elsewhere) \\
$\delta\cdot$ & delta orbital element (units of $\cdot$)\\
$x$ & radial distance from chief satellite (km)\\
$y$ & in-track distance from chief satellite (km)\\
$z$   & cross-track distance from chief satellite (km)\\
$r$  & distance from the center of the Earth to the chief satellite (km)\\
$y_p/y_a$  & in-track position when the chief is at perigee/apogee (km)\\
$p$  & parameter of ellipse $a(1-e^2)$ (km) \\
$J_2$  & coefficient from the Earth's geopotential model $1.081874\mathrm{e}{-3}$\\
$R_e$ & Earth radius $6378.137$ (km)\\
$n$ & mean motion of chief satellite \\
\end{longtable*}}

\section{Introduction}
\lettrine{R}{endezvous} and proximity operations (RPO) as well as formation flight make frequent use of the natural motions of satellites to save on the fuel required for impulsive control of satellites. As a result, it is important in the design of RPO missions and formation flying schemes to have a tool box full of natural motion behaviors which can serve as analogs for a variety of different forced motion behaviors, as these will only require the delta-v to enter into the orbit, and the minimal fuel to combat undesirable changes in the orbit due to perturbations from that point on. This paper takes the approach of Schaub and Alfriend~\cite{schaub2001J2} in using differences in mean orbital elements $[\delta a, \delta e, \delta i, \delta\Omega, \delta\omega, \delta M]$ to analyze relative motion rather than using the Hill-Clohessy-Wiltshire or Tschauner-Hempel linearized equations of motion. This choice is made to avoid the loss of periodicity one encounters when constructing periodic relative orbits with these equations and then transforming coordinates back to an inertial frame and corresponding mean orbital elements (unwanted $\delta a$ may be introduced). Further, difficulties arise from using RIC (radial, in-track, cross-track) coordinates alone to describe secular drifts due to J2, making the choice of a delta mean elements a logical one.

This paper describes relative motion using a rotating RIC coordinate system in which the origin is defined at the chief satellite's center of mass. The position of the deputy satellite is then described as $\vec{r}=x\hat{r}+y\hat{i}+z\hat{c}$, where the radial unit vector $\hat{r}$ is defined as being in the direction of position vector of the chief satellite with respect to the Earth, the cross-track unit vector $\hat{c}$ is defined as being normal to the chief's orbital plane, in the direction of the specific angular momentum, and the in-track unit vector is defined to complete the right-handed system such that $\hat{r}\times\hat{i}=\hat{c}$, making it in the general direction of the velocity, but not exactly parallel to the velocity vector unless the chief's orbit is circular.

This paper will deal only in relative motion which is periodic with the same period as the chief orbit, and thus by necessity also the deputy orbit. While, there exist relative motions that are periodic with period different from the chief, these result only from the deputy orbit having a period which is a rational multiple of the chief orbit. As such, these relative motions either exist for satellites with very different semi-major axes and thus little use for RPO or formation flight, or with periods orders of magnitudes higher than the reference orbit period. While these latter relative motions may be useful, this paper will limit itself to reference orbit periodic relative motion.

Much work has already been done in describing the construction of various relative orbit geometries employing the Tschauner-Hempel linearized equations of motion ~\cite{tschauner1965rendezvous, yamanaka2002, carter1990new, sinclair2015,bando2012graphical, sengupta2007, bae2013, alfriend2009}, and in terms of delta orbital elements in the context of reference orbits which are not highly eccentric ($e^2\sim 0$)~\cite{dang2014}. However, there is less focus in the literature on using delta orbital elements to characterize the relative motion in highly elliptical cases, though~\cite{schaub2004} did just that with their parametric equations of approximate relative motion. Further, there are many special cases for which exact analytical expressions for the relative motion may be derived for arbitrarily high eccentricities, which remain absent from the literature despite the presentation of their linearized forms. 

First, the use of a $\delta \Omega,\delta\omega,\delta M$ design space is motivated for highly elliptical reference orbits by showing the impact J2 perturbation has on relative orbits which do not obey the constraint $\delta a=\delta e=\delta i=0$. From that point, a variety of special case behaviors will be examined which one can produce within this paper's chosen design space. Combining these special cases, one can obtain many analogs to the behavior often produced by combining $\delta a,\delta e,\delta i$ in slightly elliptical cases. Approximate delta orbital elements are given in terms of desired geometric properties, with the intent of providing rules of thumb for constructing relative orbits in arbitrarily highly elliptical mission scenarios.




\section{Relative Secular Drift}
One may obtain the linearized approximations for the secular rates of change for all six mean orbital elements due to J2 perturbation using Lagrange's variation of parameters technique~\cite{vallado2001fundamentals}. It can be seen that only three of the mean orbital elements have a secular term up to first order. That is:

\begin{equation}
\label{draan}
\dot{\Omega}=-\frac{3nR_e^2J_2}{2p^2}\cos i
\end{equation}
\begin{equation}
\dot{\omega}=\frac{3nR_e^2J_2}{4p^2}(4-5\sin^2i)
\end{equation}
\begin{equation}
\label{dM}
\dot{M_0}=-\frac{3nR_e^2J_2\sqrt{1-e^2}}{4p^2}(3\sin^2i-2)
\end{equation}
Where $n$ is the mean motion, $p=a(1-e^2)$, $R_e$ is the radius of the Earth, and $J_2$ is the coefficient in the series expansion for Earth's gravitational potential due to the harmonic which represents its oblateness about the equator.

One can see that the right hand sides of equations 1-3, only depend on the classical orbital elements $a,e,i$. As per~\cite{schaub2001J2}, meaning that two orbits which share these three elements will have the same linear secular drift rates due to J2. As a result, two satellites can differ in $\Omega, \omega, M$, and still have the same secular drift rates as one another. Further, since semi-major axes are matched, two-body period will be matched, and the two satellites will exhibit a periodic relative motion. This motivates examination of periodic relative orbits resulting from these conditions.
\begin{figure}[hbt!] 
\centering
\subfloat []{\includegraphics[width=0.4\textwidth]{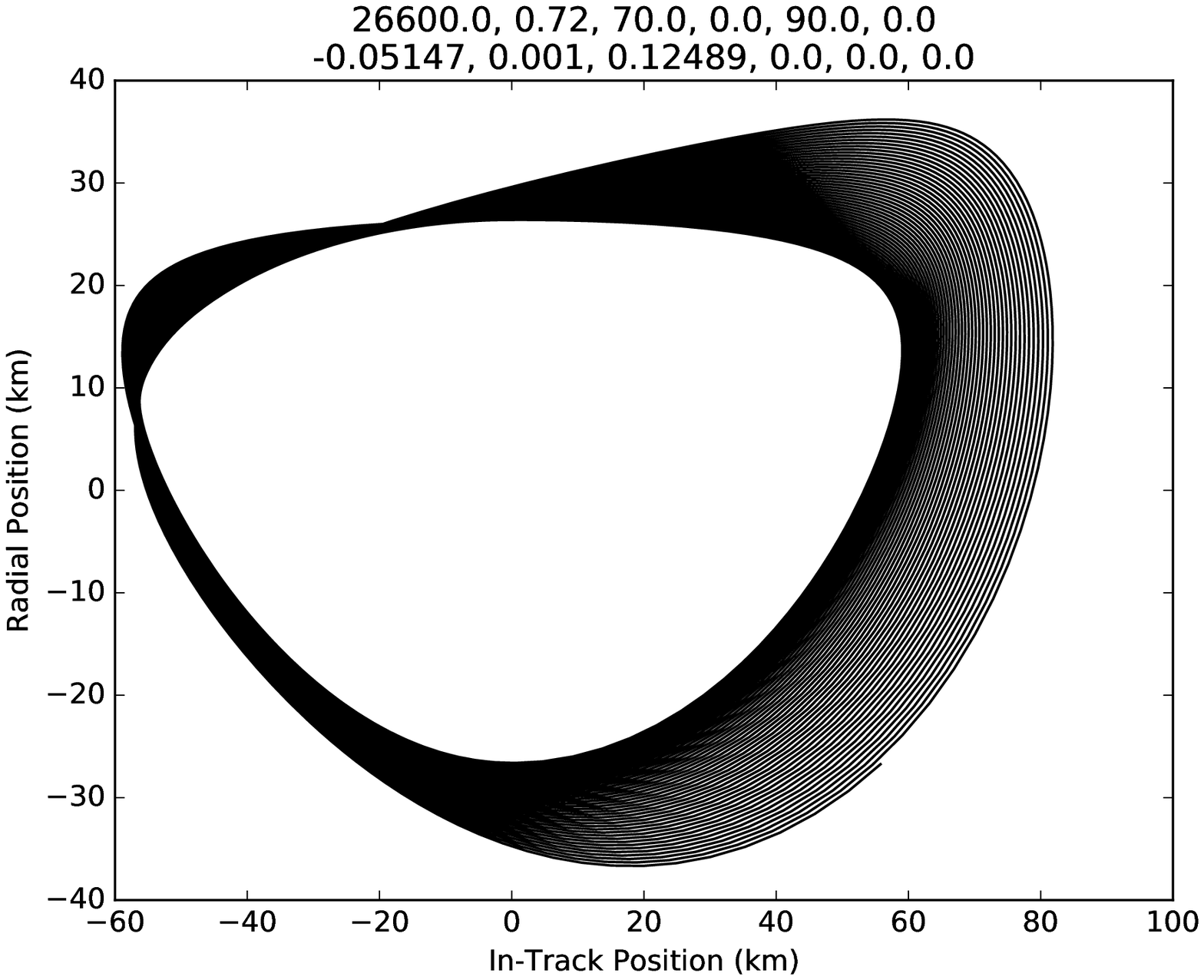}} 
\subfloat []{\includegraphics[width=0.4\textwidth]{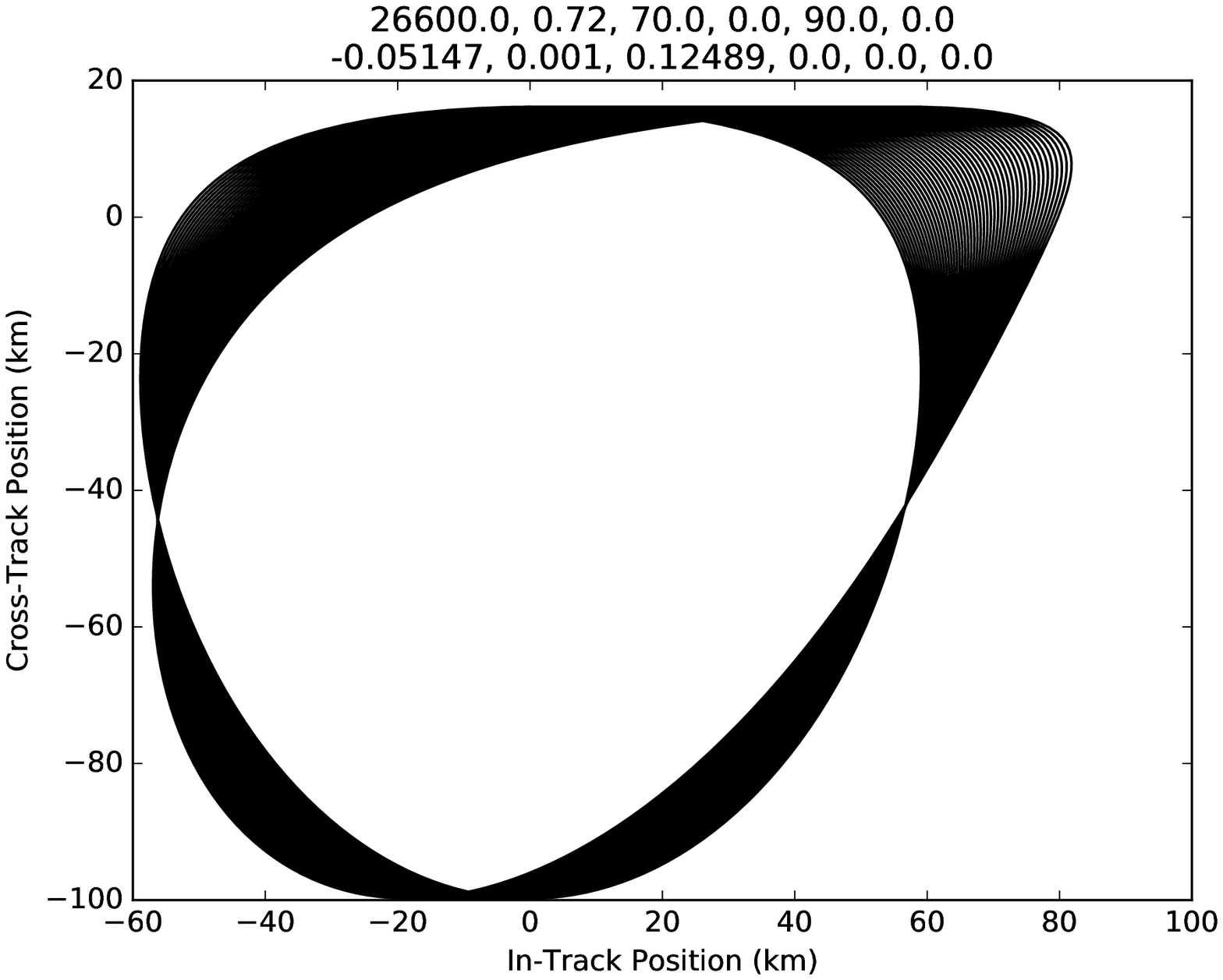}}\\
\caption{These orbits are calculated with the procedure outlined in~\cite{schaub2001J2}. Fifty periods of the chief orbit are pictured. Propagation with first order secular J2 effects.}
\label{fig:eq}
\end{figure}
\begin{figure}[hbt!] 
\centering
\subfloat []{\includegraphics[width=0.4\textwidth]{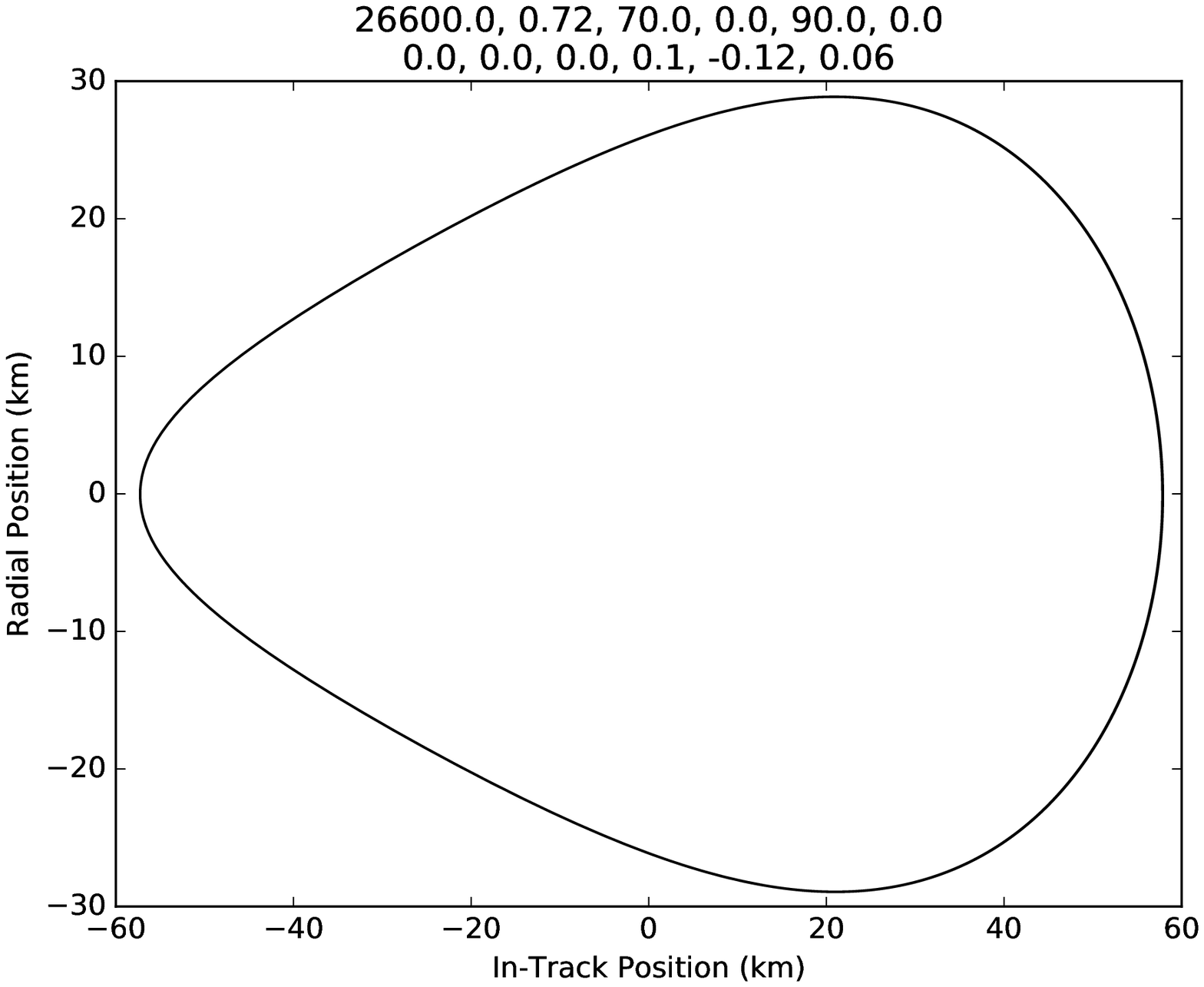}} 
\subfloat []{\includegraphics[width=0.4\textwidth]{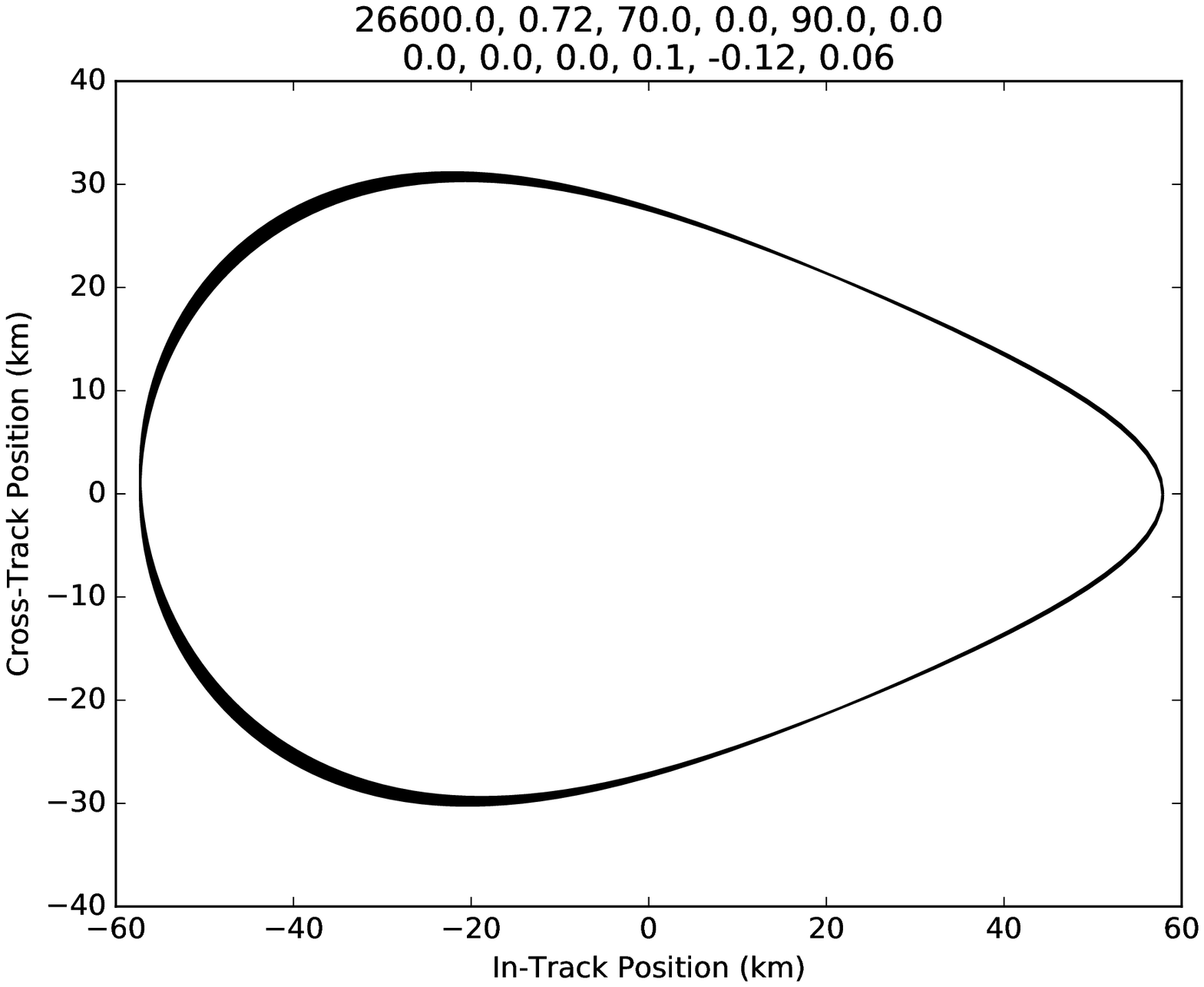}}\\
\caption{These orbits involve only $\delta\Omega,\delta\omega,\text{as well as }\delta M$, and thus do not suffer from any in-plane drifting due to linearized secular J2. Slight cross-track drift pictured is due to the chief argument of perigee drift, not the drift of any delta orbital element.}
\end{figure}
This differs from~\cite{schaub2001J2} in that one does not consider orbits with the extra degree of freedom offered by using the constraint $\dot{\delta w}=-\dot{\delta M_0}$ rather than the condition $\dot{\delta w}=0=\dot{\delta M_0}$ imposed here. While a mean latitude drift constraint provides much greater design flexibility, it can be shown for highly elliptical orbits, that allowing the perigees to drift apart causes the relative motion geometry to change shape and to grow in an unbounded fashion, in addition to the problem of a semimajor axis mismatch. So, while eliminating secular effects to J2 in a sense, drift will occur due to mismatched mean semimajor axes, and the orbit shape will change from the growth in opposite directions of $\delta M_0$ and $\delta \omega$. This is a problem unique to highly elliptical orbit contexts, and it should be noted that the $\dot{\delta w}=-\dot{\delta M_0}$ constraint is well motivated and produces J2 invariance in all but the sense of the drifting apart of the two arguments of perigee, which is not problematic unless one works in a highly elliptical context.

\section{Fundamental Periodic In-Plane Motions}
Three fundamental relative motions are described, arising from differences in the three mean orbital elements: $\Omega,\omega ,M$. The main features of interest are the relative positions of the deputy satellite as the chief satellite is at its perigee or apogee. Figures 3-10 present phase plots which include 12 points marking equal intervals in time. The triangular markers represent the passage of chief perigee, while the square markers represent passage of chief apogee. These phase plots are produced by propagating each satellite using a linearized secular J2 model individually, and then calculating the relative position at a given time using the two inertial positions at that time.

\subsection{Rectilinear Motion from \texorpdfstring{$\delta\omega$}{domega}}
\begin{figure}[hbt!]
\centering
\includegraphics[width=.5\textwidth]{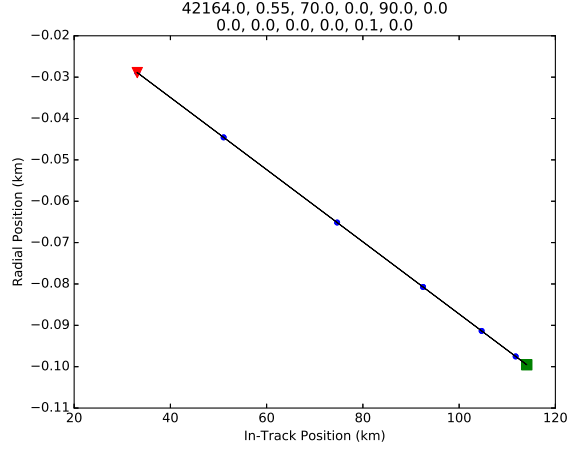}
\caption{An offset rectilinear motion induced by a $\delta\omega$}
\end{figure}

The idea of rectilinear relative motion is introduced by ~\cite{zhang2013}, in which analysis is conducted using the TH equations. This paper describes the motion as oscillation along the in-track axis, while maintaining a constant radial position of $x=0$. No comment is made as to the inertial behavior which results in this rectilinear motion; however, one can easily verify that it arises from holding all orbital elements the same between chief and deputy orbits, except for the argument of perigee $\omega$. This means that the shape and size of the two inertial orbits are the same, and that the position of the satellite within that orbit are the same ($M$ or $\nu$), but that one orbit is rotated in-plane about the Earth by $\delta\omega$.

\begin{figure}[hbt!]
\centering
\includegraphics[width=.5\textwidth]{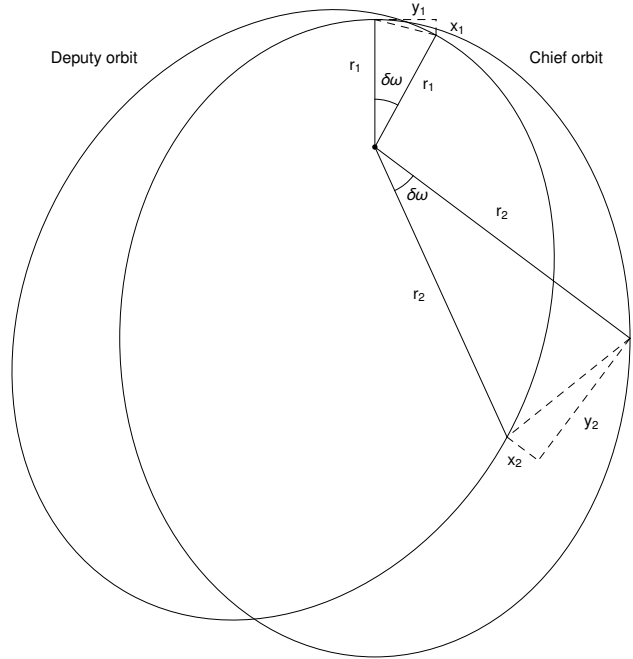}
\caption{An elliptical orbit is rotated by $\delta\omega$ about its focus (the Earth), and the relative position of two satellites are considered at two separate true anomalies. One can see that the ratio $x/y$ is a constant depending on $\delta\omega$, and that these coordinates vary based on the distance from the Earth $r$ that the two satellites share.}
\end{figure}

What results are two satellites which are always the same distance from the Earth $r=\frac{p}{1+e\cos\nu}$, but separated by a constant angle $\delta\omega$ from one another. One can then express the relative motion without any sort of approximation as:

\begin{align}
\label{rectilinear-parametric}
x&=r(\cos\delta\omega-1)\\
y&=r\sin\delta\omega
\end{align}
And as a result, the in-track extrema are:
\begin{align}
\label{rectilinear-extrema}
y_p=\frac{a(1-e^2)}{1+e}(\sin\delta\omega)=a(1-e)\sin\delta\omega\\
y_a=\frac{a(1-e^2)}{1-e}(\sin\delta\omega)=a(1+e)\sin\delta\omega
\end{align}

One can view this as the formation of an isosceles triangle with constant angle $\delta\omega$ between the two equal legs with varying magnitude $r$. As the legs on the triangle shrink and stretch as $\nu$ varies, so too does the third leg which determines the distance between the two satellite. The angle remains constant and so the motion stays on the same line which has the angle $-\delta\omega$ with the respect to the in-track axis. That is-under a Keplerian (and linearized secular J2) force model, without any approximations, a perfectly rectilinear motion occurs even for large values of $\delta\omega$; however, the motion becomes less confined to the in-track axis.

The deviation from the in-track axis is not captured by most models of relative motion, whether they be based on the TH equations or on a linearized model based off of classical orbital element differences. If one takes the linearized equations of motion from ~\cite{schaub2004,dang2014}, and uses only a nonzero value of $\delta\omega$, then the motion is predicted to stay along the intrack-axis, which is a good approximation for small $\delta\omega$, and one that will be used in later parts of this paper. Nevertheless, it is interesting that one of the fundamental solutions remains a basic geometric shape no matter the distance between chief and deputy, while the 2:1 ellipses of circular reference orbit relative motion and the offset circular relative orbit of a leader-follower formation are only approximations which work well for small chief to deputy distances.

Note also that at chief perigee, the deputy is at its closest to the chief satellite. Further, one may notice that the deputy spends most of its time far away from the chief. This is the opposite for the next fundamental motion to be studied, a fact which will become useful in designing other orbits.

\subsection{Cross-track and In-track Motion from \texorpdfstring{$\delta\Omega$}{dRAAN}}
\begin{figure}[hbt!] 
\centering
\subfloat []{\includegraphics[width=0.4\textwidth]{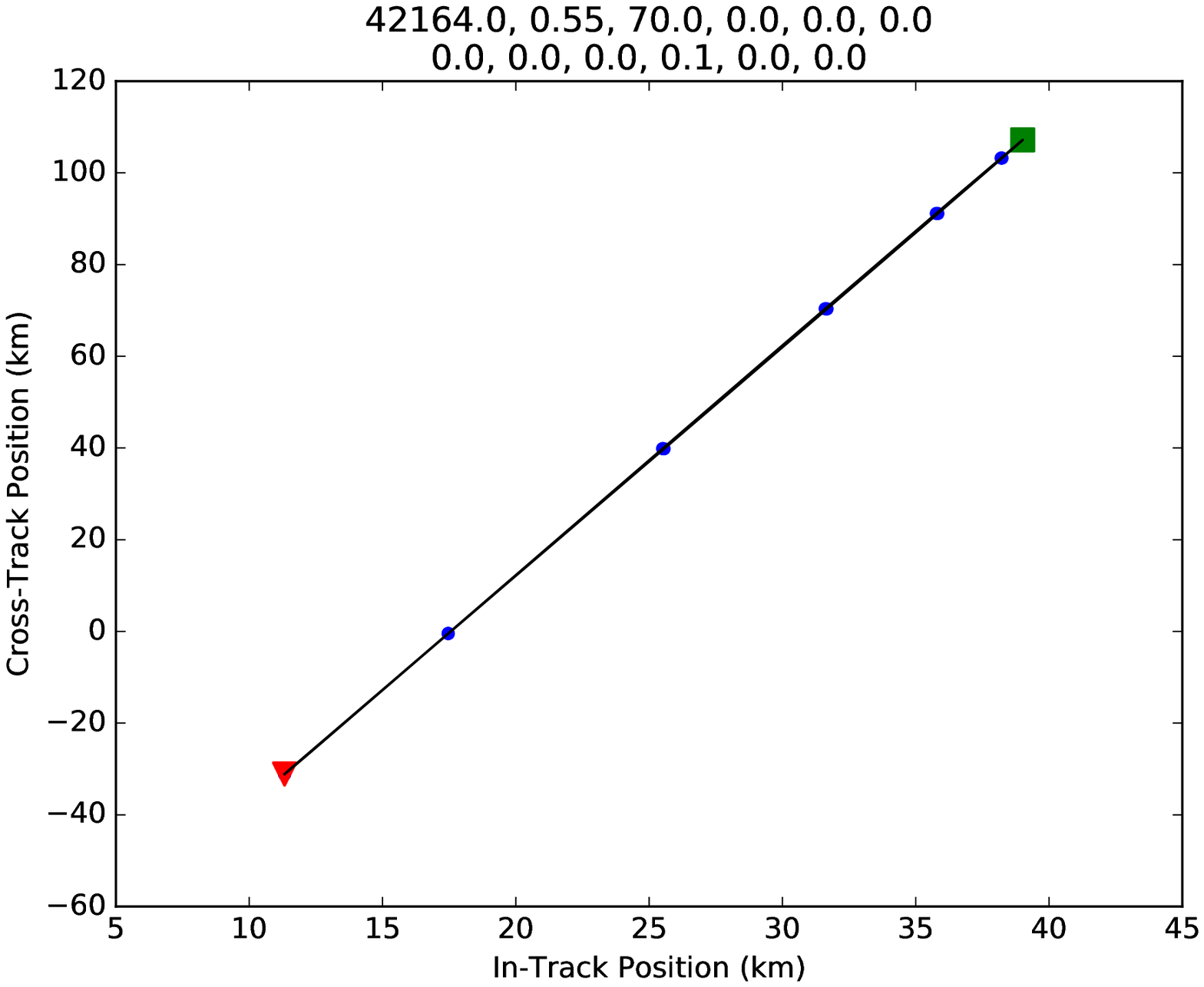}} 
\subfloat []{\includegraphics[width=0.4\textwidth]{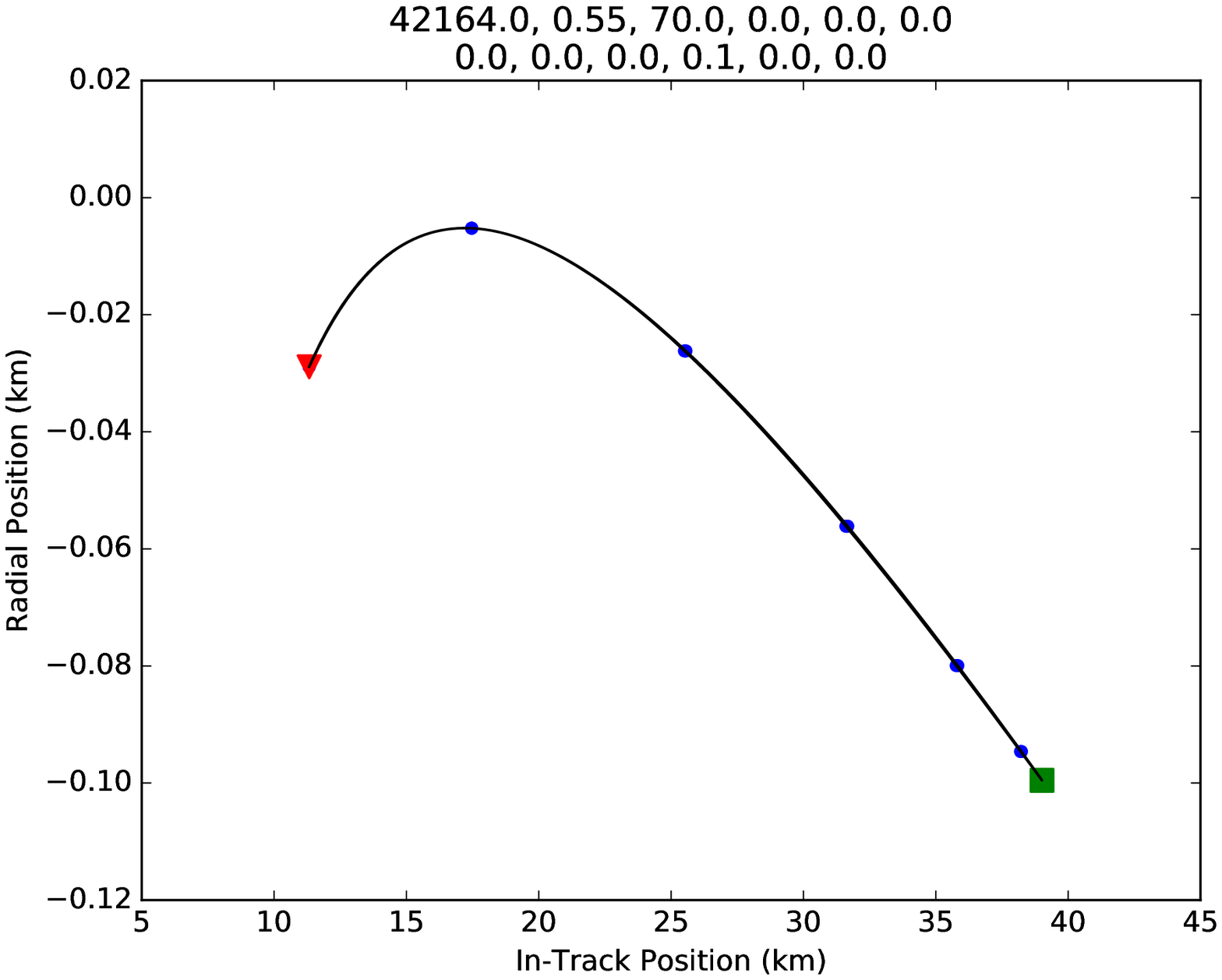}}\\
\caption{Note the difference in scale between the cross-track and in-track motion as compared with the radial motion. This is practically only a cross-track, in-track motion.}
\end{figure}
\begin{figure}[hhbt!] 
\centering
\subfloat []{\includegraphics[width=0.4\textwidth]{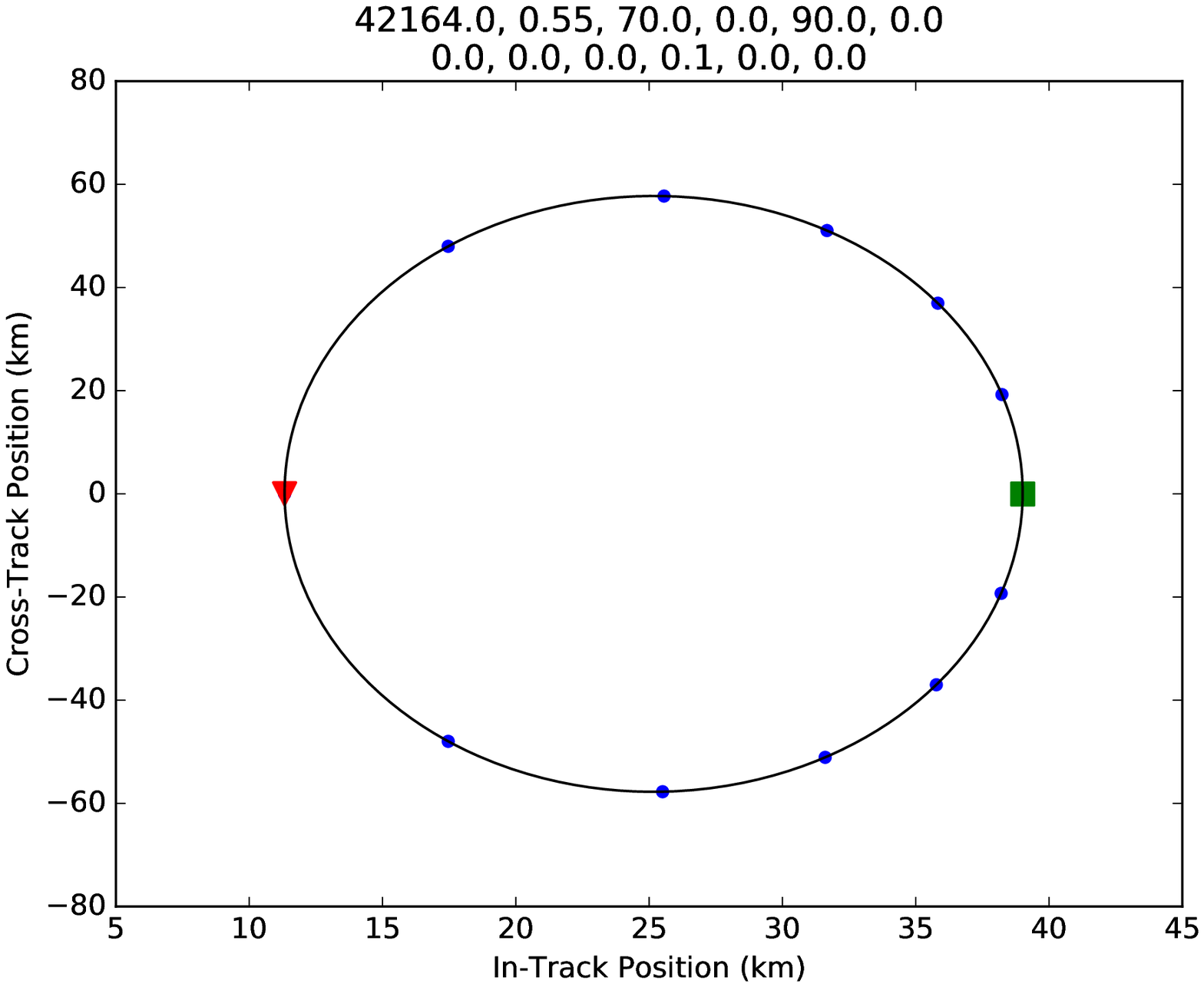}} 
\subfloat []{\includegraphics[width=0.4\textwidth]{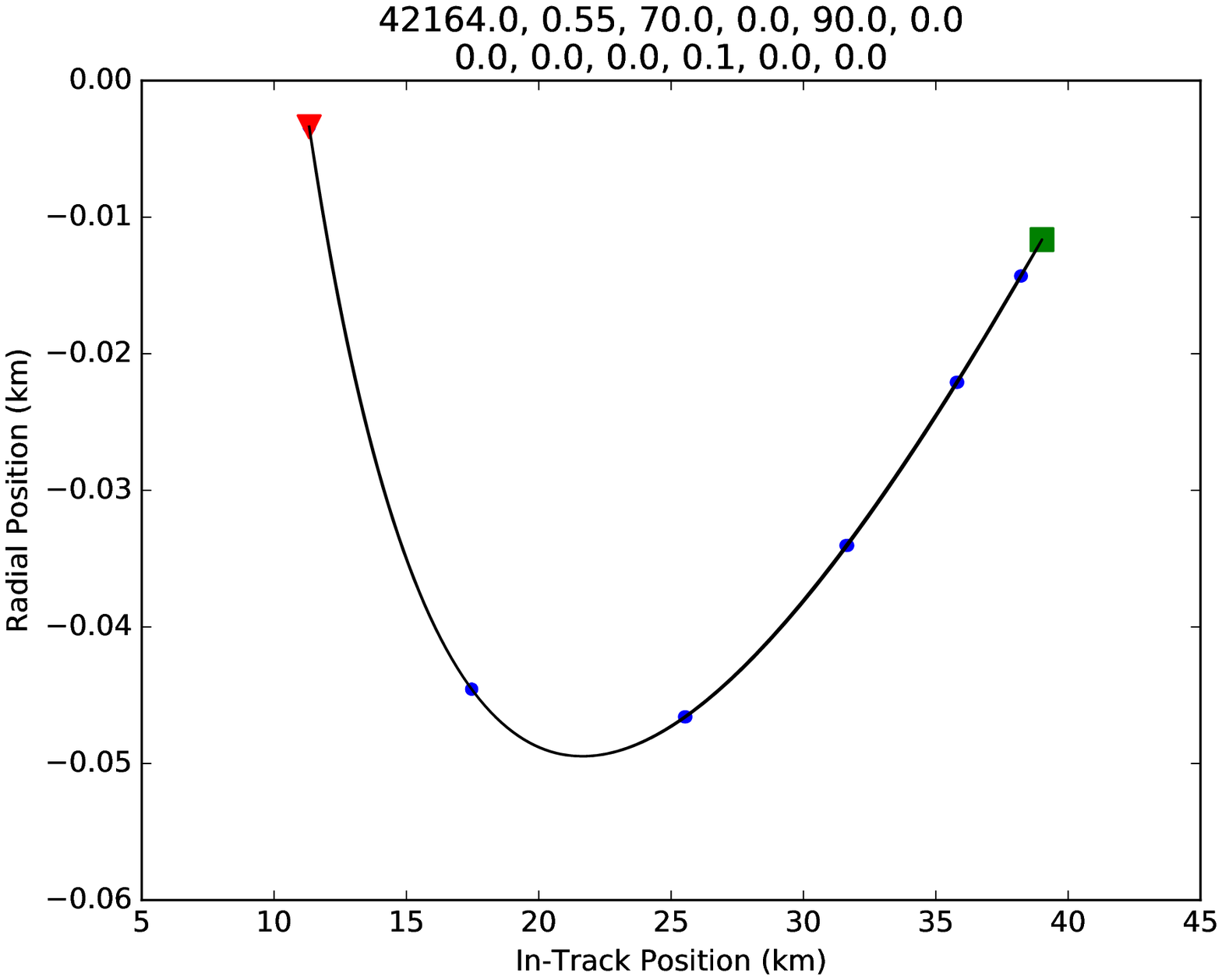}}\\
\caption{Note the difference in scale between the cross-track and in-track motion as compared with the radial motion. This is practically only a cross-track, in-track motion.}
\end{figure}
The cross-track motion induced by $\delta\Omega$ depends on the inclination and argument of perigee of the chief orbit, in addition to all of the other classical orbital elements besides the chief's $\Omega$ itself. The argument of perigee has a great influence on the size and shape of the cross-track motion, determining if it will be symmetric in the cross-track component ($\omega=90\deg$) or highly skewed as one gets close to $0\deg$. Inclination further determines the magnitude of the cross-track motion. An equatorial chief orbit has no meaning to the parameter $\delta\Omega$, and thus one cannot generate cross-track motion through this mechanism in equatorial contexts.

One may derive a parametric equation in terms of true anomaly for the relative motion induced by $\delta\Omega$ (in the case of a non-equatorial chief orbit) by expressing the deputy inertial orbit as the rotation of an arbitrary inertial chief orbit by $\delta\Omega$ about the Earth's pole, and then applying the rotation matrix which transforms the chief inertial position into a radial only vector in the RSW frame described by ~\cite{vallado2001fundamentals} to this inertial deputy position.

What results is the following description of the relative motion under Kepler's laws:

\begin{align}
x&=\frac{r}{8}[6\cos(\delta\Omega)-6+\cos(\delta\Omega-2i)-\cos(2i)+\cos(\delta\Omega+2i)-8\cos(2(\nu+\omega))\sin^2(\delta\Omega/2)\sin^2(i)]\\
y&=r\sin(\delta\Omega/2)[\cos(\delta\Omega/2-i)+\cos(\delta\Omega/2+i)+\sin(\delta\Omega/2)\sin^2(i)\sin(2(\nu+\omega))]\\
z&=-2r\sin^2(\delta\Omega/2)\sin(i)[\cos(\nu+\omega)\cot(\delta\Omega/2)-\cos(i)\sin(\nu+\omega)]
\end{align}

Upon linearization, these equations agree with~\cite{schaub2004} under the assumption of an only $\delta\Omega$ element difference. Note that this linearization removes all contribution to the radial relative motion:

\begin{align}
y&=r\cos(i)\delta\Omega\\
z&=-r\cos(\nu+\omega)\sin(i)\delta\Omega
\end{align}

And as a result, the approximate in-track extrema are:
\begin{align}
\label{raan-intrack-extrema}
y_p=\frac{a(1-e^2)}{1+e}[\cos(i)\delta\Omega]=a(1-e)\cos(i)\delta\Omega\\
y_a=\frac{a(1-e^2)}{1-e}[\cos(i)\delta\Omega]=a(1+e)\cos(i)\delta\Omega
\end{align}

Note also that any change in the chief's argument of perigee will introduce a phase shift in one component of the cross-track motion. Thus any two orbits for which $\delta\Omega$ exists will exhibit a phase shift over time in the cosine term of the cross-track motion due to perigee rotation in the chief orbit under the influence of J2. A phase shift in this one term, but not in the the other $\nu$ dependent term $r$ will cause a complete change in the cross-track motion, even changing its extrema.

In order to analyze the cross-track extrema of this motion, one takes the first derivative of the equation for the approximate cross-track motion with respect to the true anomaly. The stationary points occur at

\begin{equation}
\nu=-\sin^{-1}(e\sin\omega)-\omega
\end{equation}
Substituting this into the equation for $z$, one obtains the two extrema:

\begin{equation}
z_{stationary}\approx\pm\frac{a(1-e^2)\sqrt{1-e^2\sin^2\omega}}{1+e(\mp\cos\omega\sqrt{1-e^2\sin^2\omega}-e\sin^2\omega)}\sin(i)\delta\Omega
\end{equation}

One can see that the ratio of the two extrema is already determined by the chief orbital elements, and only the magnitude of one extremum may be controlled at a time with the design parameter $\delta\Omega$.

\subsection{In-Plane Offset Circular Motion from \texorpdfstring{$\delta M$}{dM}}

A leader-follower formation occurs when two satellites share all of the same classical orbital elements except for their mean anomaly. For small $\delta M$, the resulting shape is approximately a circle, offset along the in-track axis. This motion is analyzed using the Tschauner-Hempel equations in~\cite{sinclair2015}, or as a special case of the parametric relative motion equations in terms of delta classical elements in~\cite{schaub2004}.

Given the nonlinearity of the relationship between mean anomaly and true anomaly, the author of this paper makes no attempt to derive an exact description of the relative motion due to a mean anomaly difference between satellites. Instead, the work of~\cite{schaub2004} is presented under the assumption of a deputy with only nonzero $\delta M$, and these equations are employed in later descriptions of relative orbits composed of multiple differences in orbital elements:

\begin{equation}
x=\frac{a}{\sqrt{1-e^2}}e\sin\nu\delta M
\end{equation}
\begin{equation}
y=\frac{a}{\sqrt{1-e^2}}(1+e\cos\nu)\delta M
\end{equation}

This implies that the in-track extrema are:

\begin{align}
\label{leader-follower-extrema}
y_p=\frac{a}{\sqrt{1-e^2}}(1+e)\delta M\\
y_a=\frac{a}{\sqrt{1-e^2}}(1-e)\delta M
\end{align}

Contrary to the $\delta\omega$ and $\delta\Omega$ induced motions, the in-track minimum distance from chief occurs at chief apogee, while maximum distance is achieved at chief perigee.

\begin{figure}[hbt!]
\label{fig:leader-follower}
\centering
\includegraphics[width=.5\textwidth]{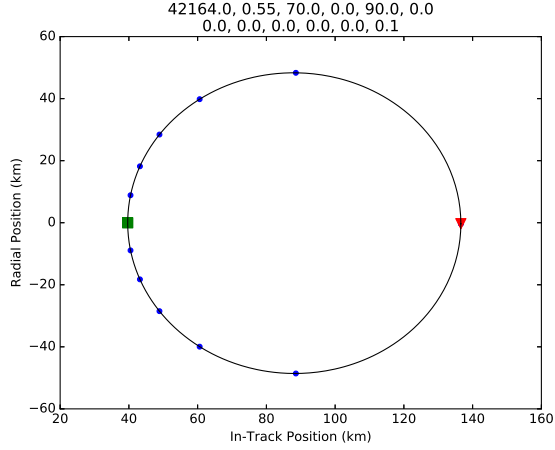}
\caption{An offset circular relative motion induced by a $\delta M$}
\end{figure}

\section{Combining \texorpdfstring{$\delta\Omega$}{dRAAN} and \texorpdfstring{$\delta\omega$}{domega}}
\begin{figure}[hhbt!] 
\centering
\subfloat []{\includegraphics[width=0.4\textwidth]{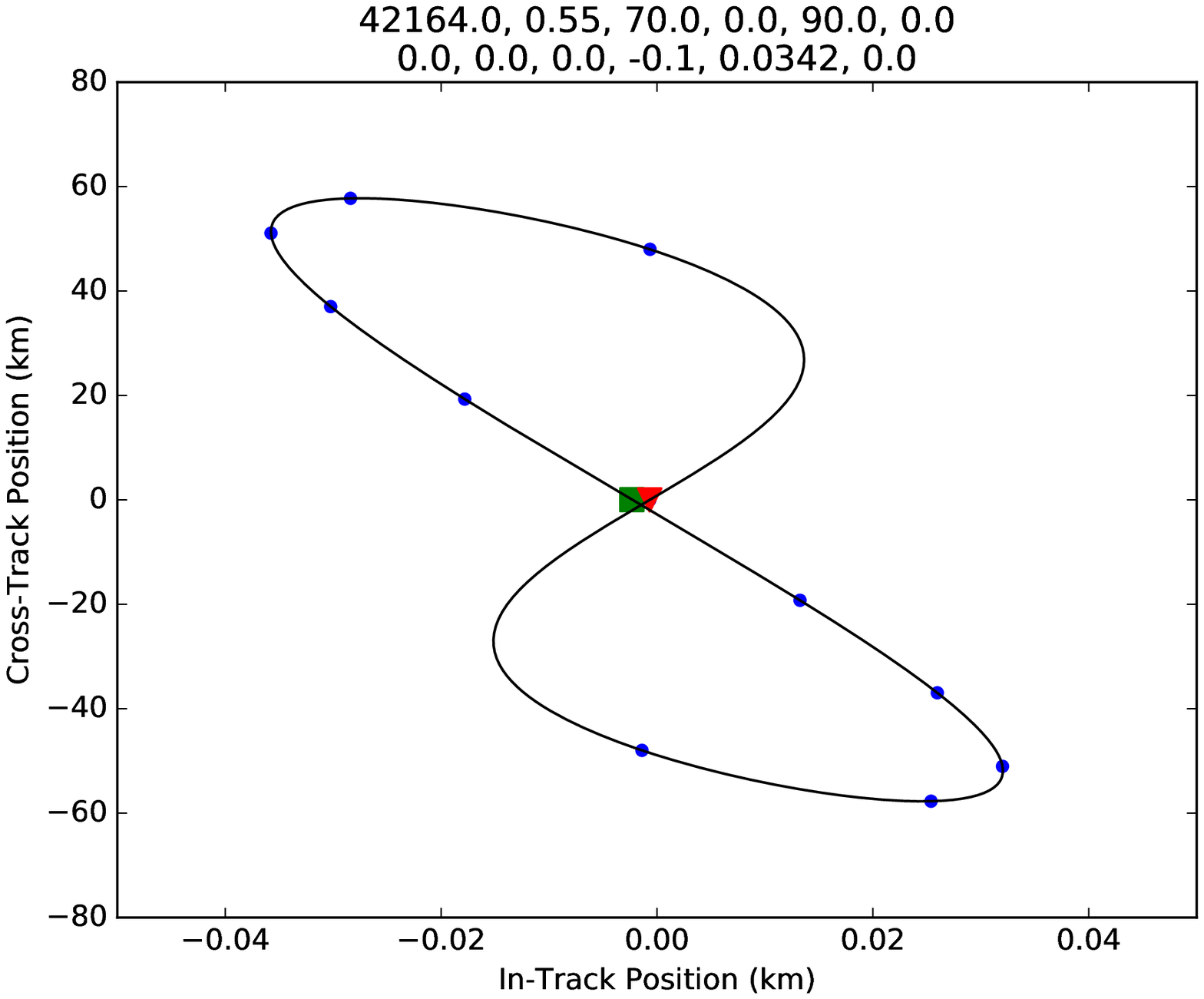}} 
\subfloat []{\includegraphics[width=0.4\textwidth]{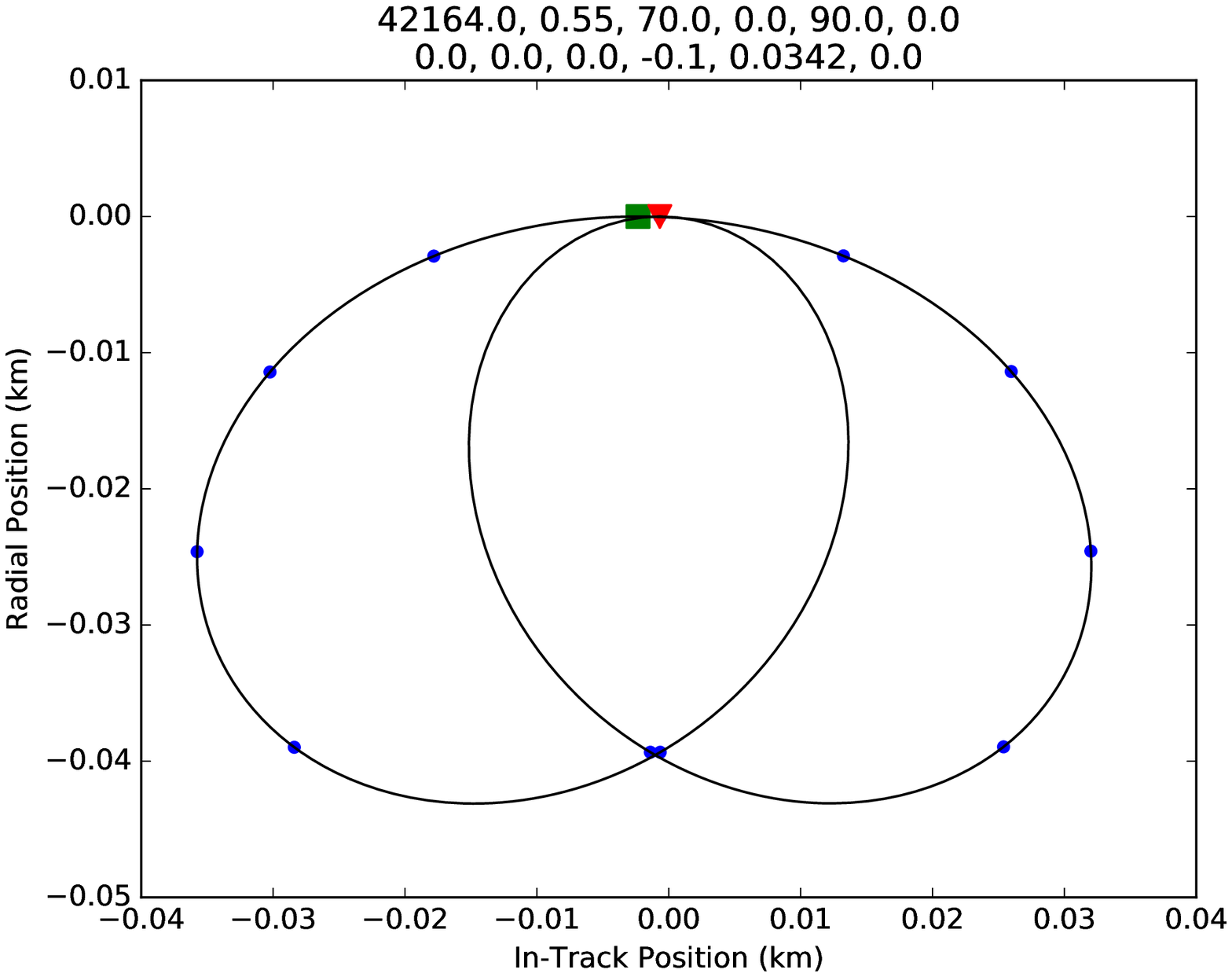}}\\
\caption{Note the difference in scale between the cross-track motion as compared with the in-track or radial motion. This is practically only a cross-track motion.}
\end{figure}
One can achieve periodic cross-track only motion by combining $\delta\Omega$ and $\delta\omega$ in specific ratios. One is, however, limited to chief orbits which are non-equatorial, and to only one type of cross-track motion (in terms of the ratio between its two extrema) determined by the chief argument of perigee. Control is possible on the amplitude of this motion.

Superimposing the in-track solutions for the motions induced by $\delta\omega$ and $\delta\Omega$, one obtains:

\begin{equation}
y=r(\cos (i)\delta\Omega+\sin(\delta\omega))
\end{equation}

Which lends the approximate constraint to cancel out the in-track motion from $\delta\Omega$
\begin{equation}
\delta\omega\approx -\cos(i)\delta\Omega
\end{equation}

\section{Combining \texorpdfstring{$\delta\omega$}{domega} and \texorpdfstring{$\delta M$}{dM}}
One can achieve periodic relative orbit geometries which contain the chief satellite by combining opposite signs of $\delta M$ and $\delta\omega$ in the correct ratios. Further, one can produce offset, non-convex geometries by combining the correct ratios of same-sign $\delta M$ and $\delta\omega$.

This behavior can be seen as stemming from a superposition principle of relative positions of the two individual fundamental periodic relative motion. This is, in effect, inducing another linearization in which one assumes that the coordinate systems of the chief and an imaginary chief at the position induced by only performing one of the fundamental motions (just the $\delta M$ or just the $\delta\omega$) are oriented closely enough that one can make the approximation that they are the oriented in the same way.

With this sort of linearizing assumption, one can create orbit geometries with specified in-track approximate extrema by using the superpositions of the two apogee fundamental motion relative positions and of the two apogee fundamental motion relative positions. The ratios of the leader-follower perigee and apogee chief passages, and the same ratio for the rectilinear fundamental motion are the following

\begin{align}
\frac{y_{pl}}{y_{al}}&\approx\frac{1+e}{1-e}\\
\frac{y_{pr}}{y_{ar}}&=\frac{1-e}{1+e}
\end{align}

Using these relationships, and linearized forms of the in-track extrema for both fundamental motions one obtains the following linear system relating in-track perigee and apogee passage positions 
(the in-track extrema when the shape is convex and the $\delta M$ and $\delta\omega$ are of opposite signs) to $\delta M$ and $\delta\omega$. 

\begin{equation}
\begin{bmatrix}
y_p\\y_a
\end{bmatrix}
\approx a
\begin{bmatrix}
1+e & 1-e\\
1-e & 1+e
\end{bmatrix}
\begin{bmatrix}
\frac{\delta\ M}{\sqrt{1-e^2}}\\
\delta\omega
\end{bmatrix}
\end{equation}

Since the column vectors of this matrix form a basis for $\mathbb{R}^2$, one can achieve a relative orbit with any perigee and apogee in-track positions using just $\delta M$ and $\delta\omega$ (the same is true of $\delta M$ and $\delta\Omega$, but not $\delta\omega$ and $\delta\Omega$).




\subsection{Centered Asymmetric Relative Orbit}
Choose $y_p=-y_a$, and what results is an asymmetric orbit with chief located equidistant between the two in-track extrema. This implies the condition

\begin{equation}
\frac{\delta M}{\sqrt{1-e^2}}=-\delta\omega
\end{equation}

This case is examined by~\cite{sengupta2007} in terms of the Tschauner-Hempel equations as a bias correction in a spatial variable.

\begin{figure}[hbt!]
\centering
\includegraphics[width=.5\textwidth]{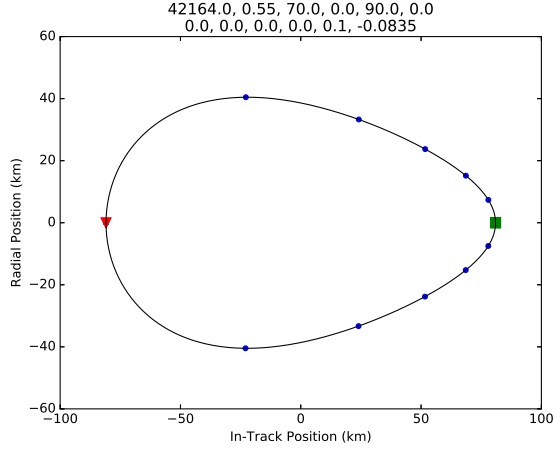}
\caption{A relative orbit induced by a specific ratio of $\delta\omega$ and $\delta M$ to center the chief in-track-wise.}
\end{figure}

\subsection{Quadrant-Time-Centered Relative Orbit}
Using the correct combination of $\delta\omega$ and $\delta M$, one may produce the only in-plane relative orbit which has the deputy spending equal amounts of time in each quadrant relative to the chief ($\delta e$ induces a relative orbit which spends equal times on either in-track side of the chief, but much more time above or below the chief depending on the sign of $\delta e$).

Examining the true anomaly of the chief satellite at a quarter period after perigee (naming this quantity $\nu_q$, and calculating it by solving Kepler's equation for $M=\pi/2$), one desires to set the in-track contributions of $\delta\omega$ and $\delta M$ equal but opposite. This leads to the constraint:

\begin{equation}
\delta M=-\frac{(1-e^2)^{3/2}}{(1+e\cos\nu_q)^2}\delta\omega
\end{equation}

This case is examined by~\cite{sengupta2007} in terms of the Tschauner-Hempel equations as a bias correction in a temporal variable.

\begin{figure}[hbt!]
\centering
\includegraphics[width=.5\textwidth]{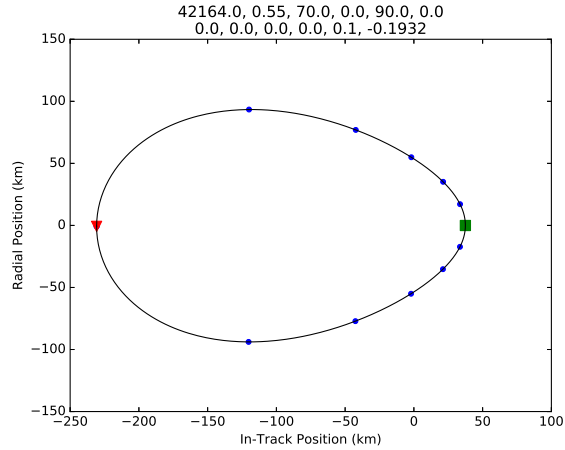}
\caption{A relative orbit induced by a specific ratio of $\delta\omega$ and $\delta M$ to center the chief in such a way that equal times are spent in each quadrant.}
\end{figure}

\subsection{Boomerang Orbit}
Reversing a sign in the centering condition presented above, one obtains a relative orbit geometry in which the perigee and apogee passage positions coincide ($y_p=y_a$). This implies the condition

\begin{equation}
\frac{\delta M}{\sqrt{1-e^2}}=\delta\omega
\end{equation}

What results is a roughly boomerang shaped orbit, which may be useful in a mission context for visiting roughly the same relative positions twice per period of the chief orbit, while exhibiting more radial motion than the rectilinear motion which has the same property. Graphs of this orbit can be seen in another paper exploring relative orbit geometry \cite{jiang2008}. This relative orbit geometry is notable for visiting roughly the same position twice per reference orbit, with one fast sweep and one slow sweep per period. Further, the deputy stays confined to a smaller in-track region than the rectilinear motion for a given distance from the chief.

\begin{figure}[hbt!]
\centering
\includegraphics[width=.5\textwidth]{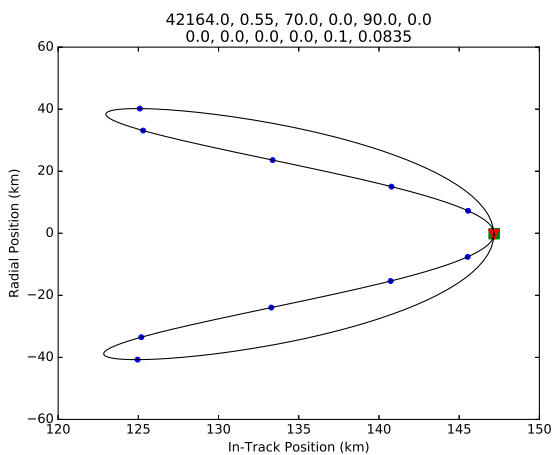}
\caption{An offset boomerang shaped orbit resulting from a specific ration of $\delta\omega$ and $\delta M$. Note the coincidence of perigee and apogee.}
\end{figure}





\section{Conclusion}
Parametric equations of relative motion, exact for Keplerian motion, and also for linearized secular J2, not presented before without linearizations were constructed using elementary means for the special cases of $\delta\Omega$ and $\delta\omega$ induced motion. These equations were then employed using a superposition principle to generate a variety of desirable orbit geometries.

This superposition method for constructing periodic relative motions consisting of $\delta\Omega$, $\delta\omega$, and $\delta M$, is useful primarily because it allows for the construction of special cases of orbits which are periodic and J2-invariant, and even further have a variety of useful properties for real life mission settings. These orbits differ from those presented in Schaub and Alfriend's paper~\cite{schaub2001J2}, because they do not have mismatched mean orbital element semimajor axes or relative argument of perigee growth, and thus remain periodic and J2-invariant even in highly elliptical contexts. One can superimpose any of the in-plane orbits mentioned in the paper with the cross-track only motion, to keep the in-plane projection as is, but induce the given cross-track motion as well.

The main disadvantage of this method is that it is difficult to find single impulses to put a deputy satellite into a given relative orbit, and instead one will have to rely on two impulses to achieve a desired behavior. This poses the natural question: which periodic J2-invariant orbits can a satellite enter from a given relative position when the chief satellite is at a given position in its own orbit? Determining those relative orbits for which one only requires a single burn to enter is a problem for future exploration with this method. Another disadvantage stems from the constraint on the types of cross-track motions one can produce from $\delta\Omega$ alone.

This work on very special cases of relative orbit geometries is just that-special case-and should be compared with the more general works of ~\cite{schaub2004,schaub2001J2,sinclair2015,sengupta2007} when designing relative orbits.



\section*{Funding Sources}

The author would like to thank the \underline{CH} and Helen Jones Foundation for their support.

\section*{Acknowledgments}
The author would like to thank Dr. Terry Alfriend, Dr. Amin Rahman, Dr. K.R. Long, Dr. George E. Pollock, Dr. Josue Munoz, Dr. Andrew Rogers, Matt Schmitt, and Izze Penelope Moore for informative discussions.
\bibliography{main}

\begin{thebibliography}{14}
\newcommand{\enquote}[1]{``#1''}
\providecommand{\natexlab}[1]{#1}
\providecommand{\url}[1]{\texttt{#1}}
\providecommand{\urlprefix}{URL }
\expandafter\ifx\csname urlstyle\endcsname\relax
  \providecommand{\doi}[1]{doi:\discretionary{}{}{}#1}\else
  \providecommand{\doi}{doi:\discretionary{}{}{}\begingroup
  \urlstyle{rm}\Url}\fi

\bibitem[{Schaub and Alfriend(2001)}]{schaub2001J2}
Schaub, H., and Alfriend, K.~T., \enquote{J 2 invariant relative orbits for
  spacecraft formations,} \emph{Celestial Mechanics and Dynamical Astronomy},
  Vol.~79, No.~2, 2001, pp. 77--95.

\bibitem[{Tschauner and Hempel(1965)}]{tschauner1965rendezvous}
Tschauner, J., and Hempel, P., \enquote{Rendezvous zu einem in elliptischer
  Bahn umlaufenden Ziel,} \emph{Astronautica Acta}, Vol.~11, No.~2, 1965, pp.
  104--+.

\bibitem[{Yamanaka and Ankersen(2002)}]{yamanaka2002}
Yamanaka, K., and Ankersen, F., \enquote{New state transition matrix for
  relative motion on an arbitrary elliptical orbit,} \emph{Journal of guidance,
  control, and dynamics}, Vol.~25, No.~1, 2002, pp. 60--66.

\bibitem[{Carter(1990)}]{carter1990new}
Carter, T.~E., \enquote{New form for the optimal rendezvous equations near a
  Keplerian orbit,} \emph{Journal of Guidance, Control, and Dynamics}, Vol.~13,
  No.~1, 1990, pp. 183--186.

\bibitem[{Sinclair et~al.(2015)Sinclair, Sherrill, and Lovell}]{sinclair2015}
Sinclair, A.~J., Sherrill, R.~E., and Lovell, T.~A., \enquote{Geometric
  interpretation of the Tschauner--Hempel solutions for satellite relative
  motion,} \emph{Advances in Space Research}, Vol.~55, No.~9, 2015, pp.
  2268--2279.

\bibitem[{Bando and Ichikawa(2012)}]{bando2012graphical}
Bando, M., and Ichikawa, A., \enquote{Graphical generation of periodic orbits
  of Tschauner-Hempel equations,} \emph{Journal of Guidance, Control, and
  Dynamics}, Vol.~35, No.~3, 2012, pp. 1002--1007.

\bibitem[{Sengupta and Vadali(2007)}]{sengupta2007}
Sengupta, P., and Vadali, S.~R., \enquote{Relative motion and the geometry of
  formations in Keplerian elliptic orbits with arbitrary eccentricity,}
  \emph{Journal of Guidance, Control, and Dynamics}, Vol.~30, No.~4, 2007, pp.
  953--964.

\bibitem[{Bae and Kim(2013)}]{bae2013}
Bae, J., and Kim, Y., \enquote{Revisiting the general periodic relative motion
  in elliptic reference orbits,} \emph{Acta Astronautica}, Vol.~85, 2013, pp.
  100--112.

\bibitem[{Alfriend et~al.(2009)Alfriend, Vadali, Gurfil, How, and
  Breger}]{alfriend2009}
Alfriend, K., Vadali, S.~R., Gurfil, P., How, J., and Breger, L.,
  \emph{Spacecraft formation flying: Dynamics, control and navigation}, Vol.~2,
  Elsevier, 2009.

\bibitem[{Dang et~al.(2014)Dang, Wang, and Zhang}]{dang2014}
Dang, Z., Wang, Z., and Zhang, Y., \enquote{Modeling and analysis of the bounds
  of periodical satellite relative motion,} \emph{Journal of Guidance, Control,
  and Dynamics}, Vol.~37, No.~6, 2014, pp. 1984--1998.

\bibitem[{Schaub(2004)}]{schaub2004}
Schaub, H., \enquote{Relative orbit geometry through classical orbit element
  differences,} \emph{Journal of Guidance, Control, and Dynamics}, Vol.~27,
  No.~5, 2004, pp. 839--848.

\bibitem[{Vallado(2001)}]{vallado2001fundamentals}
Vallado, D.~A., \emph{Fundamentals of astrodynamics and applications}, Vol.~12,
  Springer Science \& Business Media, 2001.

\bibitem[{Zhang et~al.(2013)Zhang, Zhou, Sun, and Cao}]{zhang2013}
Zhang, G., Zhou, D., Sun, Z., and Cao, X., \enquote{Optimal periodic relative
  orbit and rectilinear relative orbits with eccentric reference orbits,}
  \emph{Celestial Mechanics and Dynamical Astronomy}, Vol. 117, No.~2, 2013,
  pp. 137--148.

\bibitem[{Jiang et~al.(2008)Jiang, Li, Baoyin, and Gao}]{jiang2008}
Jiang, F., Li, J., Baoyin, H., and Gao, Y., \enquote{Study on of spacecraft
  formations in elliptical reference orbits,} \emph{Journal of guidance,
  control, and dynamics}, Vol.~31, No.~1, 2008, pp. 123--134.

\end{thebibliography}

\end{document}